\input{epsf}

\documentclass[11pt,letterpaper]{article}

\usepackage[hang,perpage]{footmisc}

\usepackage{amsmath}
\usepackage{amsthm}
\usepackage{amssymb}
\usepackage{stmaryrd}
\usepackage{graphicx}
\usepackage{latexsym}
\usepackage{txfonts}
\usepackage[varumlaut]{yfonts}
\usepackage[usenames]{color}
\usepackage{textcomp}
\usepackage{makeidx}
\usepackage{tocbibind}
\usepackage[bookmarks=true,backref=true,colorlinks=true,linkcolor=darkolivegreen,
citecolor=darkolivegreen,urlcolor=darkolivegreen]{hyperref}
\usepackage[english]{babel}
\numberwithin{equation}{subsection}
\allowdisplaybreaks

\theoremstyle{definition}
\newtheorem{ass}{Assumption}[section]
\newtheorem{theorem}[ass]{Theorem}

\def\indexname{Index of terminology}
\makeindex

\newcommand{\captionfonts}{\footnotesize}
\makeatletter  
\long\def\@makecaption#1#2{%
  \vskip\abovecaptionskip
  \sbox\@tempboxa{{\captionfonts #1: #2}}%
  \ifdim \wd\@tempboxa >\hsize
    {\captionfonts #1: #2\par}
  \else
    \hbox to\hsize{\hfil\box\@tempboxa\hfil}%
  \fi
  \vskip\belowcaptionskip}
\makeatother   
\definecolor{darkolivegreen}{rgb}{0.333333, 0.419608, 0.1843140}

\makeatletter
\def\printnotation{{%
\def\indexname{Index of notation}
\begin{theindex}
\@input{\jobname.ntn}
\end{theindex}
}}
\makeatother

\makeglossary

\begin{document}
 
\title{On a new symmetry of the solutions of  
the wave equation in the background of 
a Kerr black hole}

\author{Horst R. Beyer, Irina Craciun\thanks{Gratefully 
acknowledges financial support 
from CCT and the Department of Mathematics}
 \\ \\
 Louisiana State University (LSU) \\ 
 Center for Computation and Technology (CCT), \\
 330 Johnston Hall \\ 
\& \\
Department of Mathematics, \\
148 Lockett Hall
\\ \\
 Baton Rouge, LA 70803, USA
 \thanks{{\it E-mail:} horst@cct.lsu.edu, 
{\it Home Page:} http://www.aei.mpg.de/\texttildelow horst.}}

\date{\today}                                     

\maketitle

\pagebreak 

\begin{abstract}
\noindent
This short paper derives the constant of motion of a 
scalar field 
in the gravitational field of a Kerr black hole which is 
associated to 
a Killing tensor of that space-time. In addition, there is 
found a 
related new symmetry operator $S$ for the solutions of the 
wave equation in that background. That operator is a partial 
differential operator with 
a leading order 
time derivative of the first order that commutes with a normal form of the 
wave operator. That form
is obtained by multiplication of the wave operator from the left with 
the reciprocal of the coefficient function of its second order time 
derivative. It is shown that $S$ induces an operator 
$\hat{S}$ that commutes with
the generator of time evolution in a
formulation of the initial value problem for the wave equation in the 
setting of strongly continuous semigroups. 
\end{abstract}

\section{Introduction}
In a remarkable paper, Carter discovered that 
the Hamilton-Jacobi equation is completely separable for the Kerr space-time 
$(M,g)$.\cite{carter2} 
This result was unexpected   
since $(M,g)$
admits only a two-parameter group of isometries
which by Noether's theorem leads to two 
independent 
conserved quantities for fields in a Kerr background. 
\newline
\linebreak
The complete separability of the equation
pointed to the existence 
of another less obvious `symmetry' of $(M,g)$. 
Indeed, this property could
be traced back to the existence 
of a Killing tensor field $K$ 
which is not induced
by the isometries of $(M,g)$.\cite{walker}  
It turned out that, together with the existence of two linearly independent 
Killing vector fields  
and the existence of a further Killing-Yano tensor
field,
this fact implies also 
the complete separability of the equations 
for Spin $0,1/2,1$ and $2$ fields for 
Kerr space-time.\cite{carter3,kamran,kalnins1,kalnins2} On the 
other hand,
all these equations have been separated before without 
reference to that tensor field which might be the reason 
that some consequences of the additional symmetry 
property 
of $(M,g)$ are apparently not generally known.\cite{carter2,chandrasekhar1,
chandrasekhar2,cohen}
\newline
\linebreak
Such a consequence is that to $K$ there are also associated  
symmetry operators 
\footnote{Here and in the following, we call an operator
a symmetry operator (for the solutions of) a field equation 
if it induces
a map between classical solution spaces containing solutions
with components
belonging to a certain $C^k$-class for 
some $k \in {\mathbb{N}}^{*}$.}  
which commute with the differential
operator defining those field equations.\cite{carter4,carter3,miller,kalnins1,
kalnins2} 
For instance in the case 
of scalar fields, with respect to a local chart 
$x = (x^0,\dots,x^3)$ of $M$,
such an operator is given by \footnote{In the following, Einstein's summation 
convention is adopted whereby a repeated index  
implies summation over all
values of that index. In particular, Latin indices run from $0$ to $3$.
The symbol $\partial_{a}$ denotes the partial derivative operator 
in the $x^a$-th coordinate direction for every $a \in \{0,\dots,3\}$.}  
\begin{equation} \label{barbox}
\overline{\Box} :=
\frac{1}{\sqrt{-|g|}} \, \partial_{a} \, \sqrt{-|g|} \, K^{ab} \partial_{b} \, \, .
\end{equation}
It commutes with the wave operator 
\begin{equation*}
\Box := \frac{1}{\sqrt{-|g|}} \, \partial_{a} \, \sqrt{-|g|} 
\, g^{ab} \partial_{b} \, \, ,
\end{equation*}
i.e., 
\begin{equation*}
\left( \Box \, \overline{\Box} - \overline{\Box} \, \Box \right) u = 0 
\end{equation*}
for all $C^4$-functions $u$ defined on the domain of $x$. 
In this, $|\, \,|$ denotes the determinant function.
\newline
\linebreak
As a consequence, 
it follows by Noether's theorem the existence of 
constants of motion that are associated to classical solutions 
of the field equations of sufficiently high order of 
differentiability. Considering the 
importance of constants of motion in the analysis of 
physical systems, these 
quantities could play 
an important role in the 
stability 
discussion of fields on 
a Kerr background where they have not been used so 
far.\cite{andersson,andersson1,batic,beyer5,cardoso,damour,
detweiler,
furuhashi,finster,finster1,hartlewilkins,kamran,
khanna,konoplya,krivan1,krivan2,pressteuk,stewart,teukolsky,whiting,zouros} 
Such potential importance is also 
suggested by the 
fact that in the description 
of classical point
particles 
the constant of 
motion 
\begin{equation} \label{classicalconstantofmotion}
K(\dot{c},\dot{c}) \, \, ,
\end{equation}
which is associated to $K$ given by 
(\ref{killingtensor}), is strictly positive on 
non-trivial light-like and time-like geodesics $c$ 
were $\dot{c}$ denotes the tangent vector 
field along $c$. This is different from the
constant of motion (`energy') corresponding 
to the (`Killing-') time coordinate vector field associated 
with Boyer-Lindquist coordinates that can assume
negative values. The same 
is also true for scalar fields which considerably
complicates their stability discussion 
in a Kerr background. It might 
be possible 
to simplify the last problem by taking into account the additional 
symmetry
property of $(M,g)$.   
\newline
\linebreak  
For these reasons, in this paper, we derive the
constant of motion which is associated to $K$ 
for the case of the Klein-Gordon field in a Kerr background 
using Boyer-Lindquist 
coordinates. The derivation is an application of 
a more general discussion of 
Noether's theorem for PDEs that does not 
invoke a variational principle. 
That approach is essentially hidden 
in \cite{carter4} and is briefly sketched
in the Appendix.   
\newline
\linebreak
A drawback of the known symmetry operators  
that are associated to $K$ is that
their leading order time derivative is of the same or higher 
order than that of
the field equations.\cite{carter3,kamran,kalnins1,kalnins2}
This is different from the symmetry operators induced by Killing vector 
fields and presents an obstacle for their use 
in a rigorous functional analytic treatment of fields in a Kerr background. 
In such 
a setting \footnote{For instance, see \cite{beyer7,butzer,davies,dunford,engel,fattorini,goldstein,hillephillips,kato,krein,reed,tanabe1,yosidalec}.}, the field 
equation is reformulated as a differential equation 
of the form 
\begin{equation*}
u^{\, \prime}(t) = - G u(t)
\end{equation*} 
for all $t \in {\mathbb{R}}$. Here $u$ is a map from the real 
numbers into a 
Banach (`state'-) space $X$,${\phantom{,}}^{\prime}$ denotes the 
ordinary derivative of 
$X$-valued maps and $G$ is a densely-defined, linear operator in 
$X$ generating 
time evolution. See \cite{beyer5} for the details in the case
of the Klein-Gordon equation in the background of a Kerr black 
hole. 
\newline
\linebreak 
In such a context, it would be desirable that 
a symmetry operator would `commute' in the following 
sense with $G$. It should lead on 
an intertwining operator \footnote{Such operators are also sometimes
referred to as `constraint' operators.}  
$S$, i.e., 
a densely-defined linear operator that
 intertwines between the operators of 
the group of time evolution generated by $A$.\cite{kato5,beyer7} 
Only 
a symmetry operator whose leading 
order time derivative is
of smaller order than that of 
the field equation leads in a simple way 
to a candidate for the restriction of an intertwining 
operator to a dense domain in $X$.
\newline
\linebreak
In this paper, it will be shown that such a symmetry operator $S$
indeed 
exists for the wave equation on a Kerr background in a normalized 
form. 
That form
is obtained from the original geometric form by 
multiplication 
with the reciprocal of the coefficient of the leading 
order time derivative. Indeed, it is the last form which 
is the starting point for the reformulation of the wave equation 
in a functional analytic setting.\cite{beyer5} 
As a consequence, differently to other known symmetry operators 
connected to $K$, that symmetry operator leads on a 
operator $\hat{S}$ which is defined on dense domain of $X$
where it commutes with the infinitesimal generator 
of time evolution $G$ given in \cite{beyer5}. Hence 
$\hat{S}$ satisfies the most basic requirement for an 
intertwining operator.
It is `time-independent' in the
the sense that it does not contain time derivatives of the 
components 
of $u$. It is a candidate for the infinitesimal generator 
of a strongly continuous group or semigroup of symmetries
on the data space $X$ which are compatible with the time 
evolution.

\section{Constants of motion for the solutions of the Klein-Gordon 
equation in a Kerr background}

\label{constantsofmotion}

In Boyer-Lindquist coordinates \footnote{In the following, 
the symbols 
$t,r,\theta,\varphi$ denote coordinate projections if not 
otherwise indicated. The domains of those projections will 
be obvious from the context.}
$(t,r,\theta,\varphi) : 
\Omega \rightarrow {\mathbb{R}}^4$
the Kerr metric $g$ is given by \cite{boyer}
\begin{align*}
& g = \left(1 - \frac{2Mr}{\Sigma}\right) dt \otimes dt + 
\frac{2Mar\sin^2\!\theta}{\Sigma} \, ( dt \otimes d\varphi 
+ d\varphi \otimes dt)
  \nonumber \\
& \qquad -
\frac{\Sigma}{\Delta} \, dr \otimes dr -
\Sigma \, d\theta \otimes d\theta  
- \left(r^2 + a^2 + \frac{2Ma^2r\sin^2\!\theta}{\Sigma}\right) 
\sin^2\!\theta  \, d\varphi \otimes d\varphi
\end{align*}
where $M$ is the mass, $a \in [0,M]$ is the rotational
parameter,
\begin{align*} 
& \Delta := r^2 - 2Mr +a^2 \, \, , \, \, \Sigma := r^2 + a^2 
\cos^2\!\theta \, \, , \, \, 
r_{+} := M + \sqrt{M^2-a^2}
\, \, , \\
& \Omega := {\mathbb{R}} \times (r_{+},\infty) \times
(-\pi,\pi) \times (0,\pi) \, \, .
\end{align*}
As a reminder on the Kerr geometry, we give the following 
basic facts. The 
coordinate vector field $\partial / \partial r$ 
becomes singular on the event horizon of Kerr space-time
given by  
$r= r_{+}$. The coordinate vector field
$\partial / \partial t$ is {\it null} on the ergosphere given by
\begin{equation*}
r = M + \sqrt{M^2 - a^2 \cos^2\!\theta} \, \, ,
\end{equation*} 
it is {\it space-like} inside and {\it time-like} outside of 
ergosphere.
Finally, 
the Kerr metric is globally hyperbolic  outside the horizon
and hence the 
Cauchy problem for the Klein-Gordon equation is well posed for 
data on any Cauchy surface. In the following, we consider only 
solutions of the Klein-Gordon equation of the form 
\begin{equation*}
u(t,r,\theta,\varphi) = \bar{u}(t,r,\theta) \, e^{im\varphi}
\end{equation*}
for $(t,r,\theta,\varphi) \in \Omega$ where $m \in {\mathbb{Z}}$.
The reduced wave operator $\Box_{\,m}$, corresponding to 
$\Box$ by the replacement of $\partial_{\varphi}$ through
multiplication by the constant function of value $im$, is given by 
\begin{align*}
& \Box_{\,m}:= \frac{1}{\Sigma} \left\{ 
\left[\frac{(r^2+a^2)^2}{\Delta} - a^2 \sin^2\!\theta 
\right] \partial^2_{t} + \frac{4imMar}{\Delta} \, 
\partial_{t} - \partial_{r} \Delta \partial_{r}  \right. \\
& \left. -
\frac{1}{\sin\!\theta} \, \partial_{\theta} \sin\!\theta \,
\partial_{\theta} - m^2 \left(\frac{a^2}{\Delta} -
\frac{1}{\sin^2\!\theta} \right)
\right\} 
\end{align*}
and the reduced Klein-Gordon equation for a field $u$ of 
mass $m_0 \geq 0$ by 
\begin{equation} \label{redkleingordon}
\Box_{\,m} \bar{u} + m_0^2 \bar{u} = 0 \, \, . 
\end{equation}
A Killing tensor $K$ of the Kerr metric is given by 
\begin{align} \label{killingtensor}
K = \, \, & a^2 \left(1 - \frac{2Mr \cos^2\!\theta}{\Sigma} \right) 
dt \otimes dt \nonumber \\
& -
\frac{a \sin^2\!\theta}{\Sigma}  
\left[r^2(r^2+a^2) + a^2 \cos^2\!\theta \, \Delta \right]
 \, ( dt \otimes d\varphi 
+ d\varphi \otimes dt) \nonumber \\
& - a^2 \cos^2\!\theta \, \frac{\Sigma}{\Delta} \, 
dr \otimes dr + r^2 \, \Sigma \, d\theta \otimes d\theta 
\nonumber \\
& + \frac{\sin^2\!\theta}{\Sigma} 
\left[r^2(r^2+a^2)^2 + a^4 \sin^2\!\theta \, \cos^2\!\theta \,
\Delta \right]
 \, d\varphi \otimes d\varphi \, \, .
\end{align}
Hence 
it follows by (\ref{barbox}) that 
the reduced operator $\overline{\Box}_{\,m}$, corresponding to 
$\overline{\Box}$ by the replacement of $\partial_{\varphi}$ through
multiplication by the constant function of value $im$, is given by 
\begin{align*}
\overline{\Box}_{\,m}:= & \, \, \frac{1}{\Sigma} \left\{ 
a^2 \left[r^2 \sin^2\!\theta + \cos^2\!\theta \, \frac{(r^2+a^2)^2}{\Delta}
\right] \partial^2_{t} \right. \\
& \left. + 2ima \left( r^2 + a^2 \cos^2\!\theta \, 
\frac{r^2+a^2}{\Delta} \right)
\partial_{t} - a^2 \cos^2\!\theta \, \partial_{r} \Delta \partial_{r}  \right. \\
& \left. +
\frac{r^2}{\sin\!\theta} \, \partial_{\theta} \sin\!\theta  \,
\partial_{\theta} - m^2 \left( \frac{r^2}{\sin^2\!\theta }
+ \frac{a^4 \cos^2\!\theta }{\Delta} \right)
\right\} \, \, .
\end{align*}
Note that this operator reduces in the special case of a Schwarzschild metric 
$a = 0$ to
\begin{equation*}
\overline{\Box}_{\,m} = - \left( 
- \frac{1}{\sin\!\theta} \, \partial_{\theta} \sin\!\theta  \,
\partial_{\theta} + \frac{m^2}{\sin^2\!\theta } \right) \, \, .
\end{equation*} 
The eigenfunctions of the 
Sturm-Liouville operator that is induced by 
the last operator are used in the standard 
reduction of the wave equation in the gravitational field of a 
Schwarzschild 
black hole to 
a sequence of decoupled wave equations in one space and 
time dimension.  
\newline
\linebreak
A hyperbolic symmetry operator used later on is given by
\begin{align*}
& \widehat{\Box}_{\,m} := a^2 \, \Box_{\,m}  - \overline{\Box}_{\,m} 
= \frac{1}{\Sigma} \left[ 
\frac{2Ma^2 r (r^2 + a^2) \sin^2\!\theta}{\Delta} \, 
\partial^2_t \right. \\
& \left. - \frac{2ima}{\Delta} \, (r^2 + a^2) (\Delta - a^2 
\sin^2\!\theta) \partial_t - a^2 \sin^2\!\theta \, 
\partial_{r} \Delta \partial_{r} - \frac{r^2+a^2}{\sin\!\theta} \, 
\partial_{\theta} \sin\!\theta  \,
\partial_{\theta} \right. \\
& \left. 
+ m^2 \left( \frac{r^2 + a^2}{\sin^2\!\theta} -
\frac{a^4 \sin^2\!\theta}{\Delta} \right)
\right] \, \, .
\end{align*}
In the following, $\Box_{\,m}$ and $\widehat{\Box}_{\,m}$ 
are assumed to have $C^2(\Omega_r,{\mathbb{C}})$ as their domains
where 
\begin{equation*}
\Omega_r := {\mathbb{R}} \times (r_{+},\infty) \times (0,\pi)
\, \, .
\end{equation*} 
Using the approach to Noether's theorem 
from Section~\ref{noetherstheorem} in the Appendix and  
after some calculation, it follows that the constant 
of motion corresponding to $\overline{\Box}$ is given
\begin{align} \label{newconstantofmotion}
& K(u) := -i 
\int_{\Omega} 
\left[j(u,\overline{\Box}\,u)\right]^{0} \, dr d\theta d\varphi \\
& = 2 \pi \int_{(r_+,\infty)\times (0,\pi)} 
\sin\!\theta \left\{ 
2ma \left[ 
\frac{(r^2+a^2) \Sigma}{\Delta} \, |\partial_{0} \bar{u}|^2 
\right. \right. \nonumber \\
& \left. \left.
\qquad \quad \, \, \, \, \qquad \, \, \,  
+ \Delta |\partial_{r} \bar{u}|^2 
+ \frac{r^2 + a^2}{\Delta} \, 
|\partial_{\theta}\bar{u}|^2
 \right. \right. \nonumber \\
& \left. \left. \quad \qquad  \, \, \, \, \qquad \, \, \,   
+ \frac{m^2 \Sigma}{ \Delta \sin^2\!\theta } \, |\bar{u}|^2 
+ \frac{m_0^2(r^2+a^2)}{\Delta} \, (\Delta - a^2 
\sin^2\!\theta) |\bar{u}^2|
\right] \right. \nonumber \\
& \left. \qquad \quad \qquad  \, \, \, \, \, \, \, \,
+ i a^2 \sin^2\!\theta \, \Delta \left[
\left( \partial_{r} 
\partial_{0}\bar{u} \right)^*  \partial_{r} \bar{u}
- (\partial_{r}\bar{u})^* \partial_{r} \partial_{0} \bar{u} \right] \right. 
\nonumber  \\
& \left. \qquad \quad \qquad \, \, \, \, \, \, \, \,
+  i \, \frac{(r^2+a^2)^2}{\Delta} 
\left[ 
 \left( \partial_{\theta} 
\partial_{0}\bar{u} \right)^*  \partial_{\theta} \bar{u}
- \left(\partial_{\theta}\bar{u}\right)^* \partial_{\theta} \partial_{0} \bar{u}
\right] \right. \nonumber  \\
& \left. \qquad \quad \qquad \, \, \, \, \, \, \, \,
+ \frac{i}{\Delta} \left[ 
m^2 \left( \frac{(r^2+a^2)^2}{\sin^2\!\theta} - a^4 \sin^2\!\theta
\right) \right. \right. \nonumber  \\
& \left. \left. \, \, \, \, \, \, \, \,
\qquad \qquad \quad 
- 2m_0^2 a^2 M r (r^2 + a^2) \sin^2\!\theta \right]
\left((\partial_0 \bar{u})^{*} \bar{u} - \bar{u}^{*} \partial_{0} \bar{u} \right) 
\right\} dr d\theta \nonumber 
\end{align}
where for simplicity 
$\bar{u} \in C^2(\Omega_r,{\mathbb{C}})$ is assumed 
as a  
solution of (\ref{redkleingordon}) such that
$u(t,\cdot)$ has a compact support contained 
in 
\begin{equation*}
\Omega_{rs} := (r_+,\infty) \times (0,\pi)
\end{equation*}
for every $t$ from some open interval of ${\mathbb{R}}$.
\newline
\linebreak
Since constants of motion can be combined into other such 
constants, 
we give also those corresponding to the Killing vector fields 
$\partial_t$ and $\partial_{\varphi}$, i.e.,
the energy $E(u)$ and the z-component of the 
angular momentum $L_z(u)$. In this again, for simplicity, 
it is assumed that
$\bar{u} \in C^2(\Omega_r,{\mathbb{C}})$ is a 
solution of (\ref{redkleingordon}) such that
$u(t,\cdot)$ has a compact support contained 
in $\Omega_{rs}$ for every $t$ from some open interval of ${\mathbb{R}}$.
\begin{align*}
& E(u) :=  - \frac{1}{2} 
\int_{\Omega} 
\left[j(u,\partial_0 u)\right]^{0} \, dr d\theta d\varphi  \\
& = \pi \int_{(r_+,\infty)\times (0,\pi)} 
\sin\!\theta \left\{ 
\frac{(r^2+a^2) \Sigma + 2 M a^2 r \sin^2\!\theta}{\Delta}
\, |\partial_0 \bar{u}|^2 + \Delta |\partial_r \bar{u}|^2 \right. \\
& \left. \qquad \, \, + 
|\partial_{\theta} \bar{u}|^2 + m_0^2 \Sigma |\bar{u}|^2 + 
\frac{m^2}{\Delta \sin^2\!\theta} \,
(\Delta - a^2 \sin^2\!\theta) |\bar{u}|^2 
\right\} dr d\theta
\end{align*}
and 
\begin{align*}
& L_z(u) :=  - 
\int_{\Omega} 
\left[j(u,\partial_\varphi u)\right]^{0} \, dr d\theta d\varphi \\
& = \pi  \int_{(r_+,\infty)\times (0,\pi)} 
\frac{\sin\!\theta}{\Delta} 
\left\{ 
i m \left[ 
(r^2+a^2) \Sigma + 2 M a^2 r \sin^2\!\theta
\right] \left[ 
(\partial_0 \bar{u})^{*} \bar{u} \right. \right. \\
& \left. \left. \qquad \, \,  - \bar{u}^{*} \partial_0 \bar{u}
\right] + 4 m^2 M a r |\bar{u}|^2 
\right\} dr d\theta \, \, .
\end{align*}

\section{A new symmetry of the solutions of the wave equation in the 
background of Kerr black hole}

For the derivation of such an operator, we first give a proof 
of the fact that $\Box_{\,m}$ and $\overline{\Box}_{\,m}$ commute. 
This fact is a consequence of a more general result in 
\cite{carter4}. For the special case considered in this paper, 
a simpler direct proof is possible. Also, a similar proof 
will be given for the new symmetry operator.

\begin{theorem}
Let $u \in C^4(\Omega_{r},{\mathbb{C}})$. Then 
\begin{equation*}
\Box_{\,m} \, \overline{\Box}_{\,m} u = \overline{\Box}_{\,m} \, 
\Box_{\,m} u \, \, .
\end{equation*} 
\end{theorem} 

\begin{proof} 
Following \cite{kalnins1}, 
we define for every $n \in {\mathbb{Z}}$ the linear operators 
$D_n, D^{+}_n, L_n, L^{+}_n :  C^1(\Omega_{r},{\mathbb{C}})
\rightarrow  C(\Omega_{r},{\mathbb{C}})$ by 
\begin{align*}
& D_n v := \left\{ \partial_r + \frac{1}{\Delta} \left[ 
(r^2 + a^2) \partial_t + ima  
\right] + \frac{2n(r-M)}{\Delta} \right\} v \, \, ,  \\
& D^+_n v := \left\{ \partial_r - \frac{1}{\Delta} \left[ 
(r^2 + a^2) \partial_t + ima  
\right] + \frac{2n(r-M)}{\Delta} \right\} v \, \, , \\
& L_n v := \left[ \partial_{\theta} + \left(
- i a \sin\!\theta \, \partial_{t} + \frac{m}{\sin\!\theta}
\right) + n \, \frac{\cos\!\theta}{\sin\!\theta} \right] v \, \, , \\
& L^+_n v := \left[ \partial_{\theta} - \left( 
- i a \sin\!\theta \, \partial_{t} + \frac{m}{\sin\!\theta}
\right) + n \, \frac{\cos\!\theta}{\sin\!\theta} \right] v 
\end{align*}
for every 
$v \in C^1(\Omega_{r},{\mathbb{C}})$. In addition, 
we define 
$D^2,L^2, L^{+}_n :  C^2(\Omega_{r},{\mathbb{C}})
\rightarrow  C(\Omega_{r},{\mathbb{C}})$ by 
\begin{align*}
& D^2v := (D_1 D^+_0 + D^+_1 D_0)v \\
& = 
\frac{2}{\Delta} \, \partial_{r} \Delta \partial_{r}
- \frac{2}{\Delta^2} \left[ 
(r^2+a^2)^2 \partial_t^2 + 2ima (r^2+a^2)\partial_t - m^2a^2
\right] v
\, \, , \\
& L^2 v := 
(L_1 L^+_0 + L^+_1 L_0)v  \\
& = 2 \left( \frac{1}{\sin\!\theta} \, 
\partial_{\theta} \sin\!\theta \, \partial_{\theta} + 
a^2 \sin^2\!\theta \, \partial_{t}^2 + 2ima \partial_t - 
\frac{m^2}{\sin^2\!\theta} \right)v
\end{align*}
for every 
$v \in C^2(\Omega_{r},{\mathbb{C}})$.
As a consequence, it follows that 
\begin{equation} \label{repboxm}
\Box_{\,m} \, v = - \frac{1}{2 \Sigma} \left(
\Delta D^2 + L^2  
\right) v \, \, , \, \, 
\overline{\Box}_{\,m} \, v = - \frac{1}{2 \Sigma} \left( 
a^2 \cos^2\!\theta \, \Delta D^2 
- r^2 L^2  
\right) v \nonumber
\end{equation}
for all $v \in C^2(\Omega_{r},{\mathbb{C}})$
and, finally, that
{\allowdisplaybreaks
\begin{align*}
& [ \overline{\Box}_{\,m}, \Box_{\,m}] u = 
\left[ \frac{1}{2 \Sigma} \, 
a^2 \cos^2\!\theta \, \Delta D^2, 
\frac{1}{2\Sigma} \, L^2 \right] u - \left[ 
\frac{r^2}{2\Sigma} \, L^2, \frac{1}{2\Sigma} \,
 \Delta D^2
\right] u \\
& = \frac{\Delta a^2  \cos^2\!\theta}{ 2 \Sigma} \, 
D^2 \frac{1}{2\Sigma} \, L^2 \, u 
- \frac{1}{2\Sigma} \, L^2 \frac{\Delta a^2  \cos^2\!\theta}{ 2 \Sigma} \, 
D^2 u 
- \frac{r^2}{2\Sigma} \, L^2 \frac{\Delta}{2\Sigma} \, 
D^2 u \\
& \quad \, + \frac{\Delta}{2\Sigma} \, 
D^2 \frac{r^2}{2\Sigma} \, L^2 u 
= \frac{\Delta}{ 2 \Sigma} \, 
D^2 \frac{a^2  \cos^2\!\theta}{2\Sigma} L^2 \, u 
+ \frac{\Delta}{ 2 \Sigma} \, 
D^2 \frac{r^2}{ 2 \Sigma} \, L^2 u \\
& \quad \, - \frac{1}{2 \Sigma} \, L^2
\frac{\Delta a^2  \cos^2\!\theta}{ 2 \Sigma} \, D^2 u 
- \frac{1}{2 \Sigma} \, L^2 \frac{r^2 \Delta}{2\Sigma} \, D^2 u \\
& = \frac{\Delta}{4 \Sigma} \,  D^2  
L^2 u - \frac{1}{4 \Sigma} \,
L^2 \Delta D^2 u = 0 \, \, .
\end{align*}}
\end{proof}
Formally, the `removal' of 
the highest order time derivative from 
$\widehat{\Box}_{\,m}$ can be achieved as follows. For this, let $\bar{u}$ 
be a solution of 
\begin{equation} \label{waveeq}
\Box_{\,m} \bar{u} = 0 \, \, .
\end{equation}
The last equation can be solved 
for the second order time derivative of $\bar{u}$. By substitution of that 
derivative into the expression $\widehat{\Box}_{\,m} \bar{u}$, there is 
obtained a 
partial operator $S$ satisfying 
\begin{equation*}
\widehat{\Box}_{\,m} \bar{u} = S \bar{u}
\end{equation*}  
with a leading order time derivative which is of the first order. 
That operator has an obvious extension to 
$C^2(\Omega_r,{\mathbb{C}})$ 
which we denote by the same symbol $S$. It is given by
\begin{align} \label{newsymmetryoperator}
& Su = \frac{1}{(r^2+a^2) \Sigma + 2 M a^2 r \sin^2\!\theta} \, Pu \, \, , 
\\
& Pu := \left[ 
-2 i m a (r^2+a^2) \Sigma \partial_{t} - a^2 \sin^2\!\theta \, \Delta
\partial_r  \Delta \partial_r - \frac{(r^2+a^2)^2}{\sin\!\theta} \, 
\partial_{\theta} \sin\!\theta \, \partial_{\theta} \right. \nonumber \\
& \left. \qquad \quad \, +  
\frac{m^2 \Sigma}{\sin^2\!\theta} \, (\Sigma + 2 a^2 \sin^2\!\theta)
\right]u \nonumber
\end{align} 
for every $u \in C^2(\Omega_{r},{\mathbb{C}})$.
By construction, $S$ is a symmetry operator for the solutions of 
(\ref{waveeq}) since it maps $C^4$-solutions of 
(\ref{waveeq}) into $C^2$-solutions. 
On the other hand, it is not difficult to see that $S$ does not 
commute with $\widehat{\Box}_{\,m}$. The commutator 
is given in formula \ref{commutator} in the Appendix. Somewhat surprisingly, 
it turns out that $S$ 
commutes with the operator
\begin{align*}
& \frac{\Delta \Sigma}{(r^2+a^2) \Sigma + 2 M a^2 r 
\sin^2\!\theta} 
\, \Box_{\,m} = \partial^2_t + \frac{4imMar}{(r^2+a^2)  
\Sigma + 2 M a^2 r \sin^2\!\theta}
\, \partial_t \\
& + \frac{1}{(r^2+a^2) \Sigma + 2 M a^2 r \sin^2\!\theta} 
\left[ 
- \Delta \partial_r  \Delta \partial_r 
- \frac{\Delta}{\sin\!\theta} \, \partial_{\theta} \sin\!\theta \, \partial_{\theta} \right. \\
& \left. \qquad \qquad \qquad \qquad \qquad \qquad \quad 
+ \frac{m^2}{\sin^2\!\theta} (\Delta - a^2 \sin^2\!\theta)
\right] \, \, .
\end{align*}

\begin{theorem} \label{new} 
Let $u \in C^4(\Omega_{r},{\mathbb{C}})$. Then 
\begin{equation*}
\frac{\Delta \Sigma}{(r^2+a^2) \Sigma 
+ 2 M a^2 r \sin^2\!\theta} 
\, \Box_{\,m} \, S u = 
S \, \frac{\Delta \Sigma}{(r^2+a^2) \Sigma + 2 M a^2 r \sin^2\!\theta} 
\, \Box_{\,m} u \, \, .
\end{equation*} 
\end{theorem}
\begin{proof} To simplify notation in the following, we define 
\begin{equation*}
\overline{\Sigma} := 
\frac{(r^2+a^2) \Sigma + 2 M a^2 r \sin^2\!\theta}{\Delta} 
= \frac{(r^2+a^2)^2}{\Delta} - a^2 \sin^2\theta \, \, .
\end{equation*}
Then it follows from (\ref{repboxm}) that  
\begin{equation*} 
\frac{\Delta \Sigma}{(r^2+a^2) \Sigma + 2 M a^2 r \sin^2\!\theta} 
\, \Box_{\,m} \, v = - \frac{1}{2 \overline{\Sigma}}   
\left( 
\Delta D^2 + L^2 
\right) v
\end{equation*}
for every $v \in C^2(\Omega_{r},{\mathbb{C}})$.
Further, 
by a straightforward calculation, it follows that 
\begin{equation*}
Pv = - \frac{1}{2} \left[ 
a^2 \sin^2\theta \, \Delta^2 D^2 + 
(r^2 + a^2)^2 L^2
\right]v 
\end{equation*}
and hence that
\begin{align*}
& Sv = - \frac{1}{2} \,  \frac{1}{(r^2+a^2) \Sigma + 2 M a^2 r \sin^2\!\theta}
\left[ 
a^2 \sin^2\theta \, \Delta^2 D^2 + (r^2 + a^2)^2 L^2
\right]v \\
& =  - \frac{1}{2 \overline{\Sigma}}   
\left[ 
a^2 \sin^2\theta \, \Delta D^2 + 
\frac{(r^2 + a^2)^2}{\Delta} \, L^2
\right]v 
\end{align*}
for every $v \in C^2(\Omega_{r},{\mathbb{C}})$.
Finally, it follows that 
\begin{align*}
& [S, \Sigma \, \overline{\Sigma}^{\, -1} \, \Box_{\,m}]u = 
\frac{1}{4} \bigg\{
\frac{1}{\overline{\Sigma}}  
\left[ 
a^2 \sin^2\theta \, \Delta D^2 + 
\frac{(r^2 + a^2)^2}{\Delta} \, L^2
\right], \\
& \qquad \qquad \qquad \qquad \quad \, \, \frac{1}{\overline{\Sigma}} \,  
(
\Delta D^2 + L^2)
\bigg\}u \\
& 
= \frac{1}{4 \overline{\Sigma}} \left[
a^2 \sin^2\theta \, \Delta D^2 
 \frac{1}{\overline{\Sigma}} \, L^2 - 
\Delta  D^2  \frac{1}{\overline{\Sigma}} \, 
\frac{(r^2 + a^2)^2}{\Delta} \, L^2
\right]u \\
& \quad \, + \frac{1}{4 \overline{\Sigma}} \left[
\frac{(r^2 + a^2)^2}{\Delta} \, L^2
\frac{\Delta}{\overline{\Sigma}} \, D^2
- L^2 \frac{1}{\overline{\Sigma}} \, 
a^2 \sin^2\theta \, \Delta D^2
\right]u  \\
& = \frac{\Delta}{4 \overline{\Sigma}} 
\, D^2 \, 
\frac{a^2 \sin^2\theta - ((r^2+a^2)^2/\Delta)}{\overline{\Sigma}} \, 
L^2u \\
& \quad \, \, 
+ \frac{\Delta}{4 \overline{\Sigma}} \, L^2
\, \frac{((r^2+a^2)^2/\Delta) -a^2 \sin^2\theta}{\overline{\Sigma}} \, 
D^2u \\
& = - \frac{\Delta}{4 \overline{\Sigma}} 
\, D^2 \, 
L^2u + \frac{\Delta}{4 \overline{\Sigma}} \, L^2
\, D^2u = 0 \, \, .
\end{align*}
\end{proof} 
Note that the constants of motion associated to 
a solution $\bar{u}$ of the reduced wave equation
(\ref{waveeq})
and the operators $\overline{\Box}_{\,m}, \widehat{\Box}_{\,m}, S$ 
are, apart from a sign, the same since
\begin{equation*}
\overline{\Box}_{\,m} \bar{u} = - \, \widehat{\Box}_{\,m}  \bar{u}
= - S \bar{u}
\, \, . 
\end{equation*} 
In \cite{beyer5}, it was proved 
the well-posedness of the initial
value problem for the reduced Klein-Gordon in a Kerr background 
in the setting of strongly continuous semigroups of operators,  
the domain of the corresponding infinitesimal generator $G$
was investigated and its resolvent given in terms of spheroidal 
functions.
Further, the stability of the solutions was proved  
if the mass of the field exceeds a certain combination of the 
`magnetic quantum number' $m \in \mathbb{Z}$ 
relative to the rotation axis, the mass and the 
rotational parameter of the black hole. 
For details refer to 
\cite{beyer5}. In particular
for $(u,v) \in C^2_0(\Omega_{rs},{\mathbb{C}}) 
\times C_0(\Omega_{rs},{\mathbb{C}})$
\begin{equation*}
G(u,v) = (-v,(A+C)u + iBv)
\end{equation*} 
where 
\begin{align*}
& (A+C) u =  \frac{1}{(r^2+a^2) \Sigma + 2 M a^2 r \sin^2\!\theta} 
\left[ 
- \Delta \partial_r  \Delta \partial_r 
- \frac{\Delta}{\sin\!\theta} \, \partial_{\theta} \sin\!\theta \, \partial_{\theta} \right. \\
& \left. \qquad \qquad \qquad \qquad \qquad \qquad \qquad \qquad \, 
\, \, \, \, 
+ \frac{m^2}{\sin^2\!\theta} (\Delta - a^2 \sin^2\!\theta)
\right] u \, \, , \\
& Bv = \frac{4 m M a r}{(r^2+a^2) \Sigma + 2 M a^2 r \sin^2\!\theta}
\, v \, \, .
\end{align*}
It is not difficult to see that in the setting of \cite{beyer5}, 
$S$ induces the 
operator $\hat{S}$ defined by
\begin{align} \label{newsymmetryoperator1}
& \hat{S}(u,v) = \frac{1}{\Delta \overline{\Sigma}} \, 
\big(P_s u - 2 ima (r^2+a^2) \Sigma v, 
2 ima (r^2+a^2) \Sigma (A+C) u \nonumber \\ 
& \qquad \qquad \qquad 
\, \, + 
P_s v - 2ma (r^2+a^2)\Sigma B v \big)  
\end{align}
for all $u, v \in C^2_0(\Omega_{rs},{\mathbb{C}})$ 
where 
\begin{align*}
& P_s u := \left[ 
- a^2 \sin^2\!\theta \, \Delta
\partial_r  \Delta \partial_r - \frac{(r^2+a^2)^2}{\sin\!\theta} \, 
\partial_{\theta} \sin\!\theta \, \partial_{\theta} \right. \\
& \left. \qquad \quad \, \, \, +  
\frac{m^2 \Sigma}{\sin^2\!\theta} \, (\Sigma + 2 a^2 \sin^2\!\theta)
\right]u \, \, .
\end{align*}
Then the following holds
\begin{theorem}
Let $(u,v) \in 
C^4_0(\Omega_{rs},{\mathbb{C}}) 
\times 
C^2_0(\Omega_{rs},{\mathbb{C}})
$. Then 
\begin{equation} \label{intertwining}
\hat{S} G (u,v) = G \hat{S} (u,v) \, \, .
\end{equation} 
\end{theorem}

\begin{proof}
First, it follows from Theorem~\ref{new} that
\begin{align} \label{commutators}
& \left(- 2ima \left[ 
\frac{(r^2+a^2)\Sigma}{\Delta \overline{\Sigma}}, 
A+C \right] + i \left[\frac{1}{\Delta \overline{\Sigma}}
P, B \right]\right) w_1 = 0 \, \, , \nonumber \\
&  \left[\frac{1}{\Delta \overline{\Sigma}}
P, A+C \right] w_2 = 0 
\end{align}
for all $w_1 \in C^2_0(\Omega_{rs},{\mathbb{C}})$ and 
$w_2 \in C^4_0(\Omega_{rs},{\mathbb{C}})$. Further, it follows 
\begin{align*}
& \hat{S} G (u,v) \\
& = 
\frac{1}{\Delta \overline{\Sigma}}
(- 2 ima (r^2+a^2) \Sigma (A+C) u - P v + 2 ma (r^2+a^2) \Sigma B v,\\
& \, \, \quad P (A+C) u - 2 ma (r^2+a^2) \Sigma B (A+C) u \\
& \quad - 2 ima (r^2+a^2) \Sigma (A+C) v + i P B v - 2ima
(r^2+a^2) \Sigma B^2 v )  \\
& G \hat{S} (u,v) \\
& =  \frac{1}{\Delta \overline{\Sigma}}
( - 2 ima (r^2+a^2) \Sigma (A+C) u - P v + 2 ma (r^2+a^2) \Sigma B v, \\
& \, \, \quad \Delta \overline{\Sigma} (A+C) 
\frac{1}{\Delta \overline{\Sigma}} Pu 
- 2 ma (r^2+a^2) \Sigma B (A+C) u \\
& \, \quad -2ima \Delta \overline{\Sigma} (A+C)
\frac{(r^2+a^2)\Sigma}{\Delta \overline{\Sigma}} v + 
i B P v \\
& \, \quad - 2 i m a (r^2+a^2) \Sigma B^2 v )
\end{align*}
and hence by (\ref{commutators}) the 
validity of (\ref{intertwining}).
\end{proof} 

\section{Discussion and open problems}

The results open up a number of interesting new questions for future 
research. 
\newline
\linebreak
It would be interesting to know whether and, if applicable,
when the new constant of motion 
$K(u)$ \ref{newconstantofmotion} (or some combination with 
the other constants of motion) has a definite sign 
like its classical counterpart
(\ref{classicalconstantofmotion}). For instance, a definite sign 
of $K(u)$ in the case of a vanishing mass of the field would open
up the road to a simpler and independent proof of the stability of the solutions 
of the wave equation in a Kerr background.\cite{finster} In addition, 
it could lead to a sharper estimate on a possible onset of instability in 
the case of a non-vanishing mass of the field.\cite{beyer5,konoplya} 
\newline
\linebreak
Connected to the new symmetry operator $S$ 
(\ref{newsymmetryoperator})
are technical questions whether the closure of the induced 
operator
$\hat{S}$ is an intertwining operator and/or  
is the infinitesimal generator of a strongly continuous group or semigroup.
Also, with the help of the new symmetry operator $\hat{S}$, it 
should be possible to further reduce the wave equation (\ref{waveeq}), 
in a proper sense, to an equation in `one space dimension'. For this, a deeper 
study of the properties of $\hat{S}$ is necessary. In addition, it would 
be interesting to explore whether there can be made a connection of the results
here to those of \cite{whiting}. Finally, it remains the question whether 
such a symmetry operator can also be found for a non-vanishing mass of the 
scalar field.  
\newline
\linebreak
It is to expect that 
the results of the paper can be generalized 
to the Dirac, Maxwell and the linearized
Einstein's field equation around the Kerr metric.
If that turns out be the case, it would be interesting to 
investigate their potential for finding a generalization to 
fields on a Kerr background of  
the Regge-Wheeler-Zerilli-Moncrief decomposition 
of fields on a Schwarz\-schild 
background.\cite{reggewheeler,zerilli,moncrief}

\section{Appendix}

\subsection{An approach to Noether's theorem 
not invoking a variational principle}

\label{noetherstheorem}

In this Section, we briefly give an approach to the derivation of 
conservation laws from transformations mapping classical solution
spaces of a linear scalar PDE into one another. The approach 
is essentially hidden 
in \cite{carter4} 
and is not invoking a variational principle. 
The derivation of the constants of motion in 
Section~\ref{constantsofmotion}
is an application of this approach.  
\newline
\linebreak 
For this, we consider the equation

\begin{equation} \label{PDElinear}
\frac{1}{\rho} \, 
\frac{\partial }{\partial x_a} \left(
\rho \,g^{ab} \, \frac{\partial u}{\partial x_b} \right) 
+ V \, u = 0 
\end{equation}
where $n \in {\mathbb{N}}^{*}$, $\Omega$ is a non-empty 
open subset of 
${\mathbb{R}}^{n+1}$, $g^{ab} \in C^1(\Omega,{\mathbb{R}})$
for all $a,b \in \{0,\dots,n\}$ such that $g^{ab} = g^{ba}$
for all $a,b \in \{0,\dots,n\}$, $\rho \in 
C^1(\Omega,{\mathbb{R}})$
such that $\textrm{Ran} \, \rho \subset (0,\infty)$,
$V \in C(\Omega,{\mathbb{R}})$ 
and $u \in C^2(\Omega,{\mathbb{C}})$. 
\newline
\linebreak
For solutions $u,v$ of (\ref{PDElinear}), it can be derived 
a corresponding conservation law. First, it follows that 
\begin{align*}
& \frac{u^{*}}{\rho} 
\, \frac{\partial }{\partial x_a} \left(
 \rho \, g^{ab} \, \frac{\partial v}{\partial x_b} \right) + V 
\, u^{*} v = 0 \, \, , \, \, 
\frac{v}{\rho} 
\, \frac{\partial }{\partial x_a} \left(
\rho \,g^{ab} \, \frac{\partial u^{*}}{\partial x_b} \right) + V 
\, v u^{*} = 0
\end{align*}
where $\phantom{}^{*}$ denotes complex conjugation.
The difference of these equations gives 
\begin{align*}
0 & = 
\frac{u^{*}}{\rho} \, \frac{\partial }{\partial x_a} 
\left(
\rho \, g^{ab} \, \frac{\partial v}{\partial x_b} \right) -
 \frac{v}{\rho} \, \frac{\partial }{\partial x_a} \left(
\rho \, g^{ab} \, \frac{\partial u^{*}}{\partial x_b} \right) \\
& = \frac{1}{\rho} \left[
\frac{\partial }{\partial x_a} \left(
 u^{*} \rho \, g^{ab} \, \frac{\partial v}{\partial x_b} \right) -
\frac{\partial }{\partial x_a} \left(
 v \, \rho \, g^{ab} \, \frac{\partial u^{*}}{\partial x_b} \right) \right] 
\\
& =  \frac{1}{\rho} \, 
\frac{\partial }{\partial x_a} \, \rho \, g^{ab} \left( 
u^{*} \, \frac{\partial v}{\partial x_b} -
v \, \frac{\partial u^{*}}{\partial x_b} \right) 
\end{align*}
and hence 
\begin{equation*}
\frac{1}{\rho} \, \partial_{a} \, \rho \left[j(u,v)\right]^a = 0  
\end{equation*}
where the vector field $j(u,v) \in  
C^1(\Omega,{\mathbb{R}}^{n+1})$ is defined by 
\begin{equation*}
\left[ j(u,v) \right]^{a} =  
g^{ab} \left( 
u^{*} \, \frac{\partial v}{\partial x_b} -
v \, \frac{\partial u^{*}}{\partial x_b} \right)
\end{equation*}
for all $a \in \{0,\dots,n\}$. If 
$F : C^k(\Omega,{\mathbb{C}}) \rightarrow C^2(\Omega,{\mathbb{C}})$
for some $k \in \{2,3,\dots\}$  
maps $C^k$-solutions of (\ref{PDElinear})
into $C^2$-solutions of (\ref{PDElinear}), then it follows 
the constraint 
\begin{equation*}
\frac{1}{\rho} \, \partial_{a} \, \rho \left[ j(u,F(u)) \right]^a = 0    
\end{equation*} 
for every $C^k$-solution $u$ of (\ref{PDElinear}).
\newline
\linebreak
As a short example,
we consider  
the special case of a wave equation in flat space-time 
\begin{equation} \label{wavetype}
\frac{\partial^2 u}{\partial t^2} - \Delta u  + 
V u = 0   
\end{equation}
where $V \in C(\Omega,{\mathbb{R}})$ is 
such that 
\begin{equation*}
\frac{\partial V}{\partial t} = 0 
\end{equation*}
and $u$ is assumed as real-valued. If $u$ is a $C^3$-solution  
of (\ref{wavetype}),
then
\begin{equation*}
0 = \frac{\partial}{\partial t}
\left(\frac{\partial^2 u}{\partial t^2} - \Delta u  + 
V u \right) = 
\frac{\partial^2}{\partial t^2} \, \frac{\partial u}{\partial t} 
- \Delta \, \frac{\partial u}{\partial t} 
+ V \, \frac{\partial u}{\partial t} 
\end{equation*} 
and hence $\partial u / \partial t$ 
is a $C^2$-solution of (\ref{wavetype}). As a consequence,
\begin{equation*}
\frac{\partial}{\partial t}  
\left[ j\left(u,\frac{\partial u}{\partial t} \right)\right]^0 = 
- \sum_{k=1}^{n}
\frac{\partial}{\partial x_k} 
\left[ j\left(u,\frac{\partial u}{\partial t} \right) \right]^k
\, \, .  
\end{equation*}
Further,
\begin{align*}
& \left[ j\left(u,\frac{\partial u}{\partial t}\right)\right]^0 =   
u \, \frac{\partial^2 u}{\partial t^2} -
\left(\frac{\partial u}{\partial t}\right)^2 =
u \cdot \left( \Delta u - V u \right) - 
\left(\frac{\partial u}{\partial t}\right)^2 \\
& = u \Delta u 
- \left[\left(\frac{\partial u}{\partial t}\right)^2 + V u^2 
\right] 
= 
\nabla \left( u \nabla u 
\right) - \left[\left(\frac{\partial u}{\partial t}\right)^2 + 
|\nabla u|^2 + V u^2 \right] \\
& = 
\nabla \left( u \nabla u \right) - 2 \, \epsilon(u)
\end{align*}
where 
\begin{equation*}
\epsilon(u) := 
\frac{1}{2} \,
\left[\left(\frac{\partial u}{\partial t}\right)^2 + 
|\nabla u|^2 + V u^2 \right] 
\end{equation*}
and $\nabla$ denotes the gradient in the spatial coordinates.
In addition,
\begin{align*}
\left[ j\left(u,\frac{\partial u}{\partial t}\right)\right]^k 
=
- \left( 
u \, 
\frac{\partial^2 u}{\partial x_k \partial t} -
\frac{\partial u}{\partial t} \, 
\frac{\partial u}{\partial x_k} \right) \, \, .
\end{align*}
Hence it follows that 
\begin{equation*}
\frac{\partial }{\partial t} \left[
\nabla \left( u \nabla u \right) - 2 \, \epsilon(u) \right] = 
\nabla \left( 
u \, \nabla \, \frac{\partial u}{\partial t} -
\frac{\partial u}{\partial t} \, \nabla u \right) \, \, .
\end{equation*}
The last implies that 
\begin{equation*}
\frac{\partial}{\partial t} \left( - 2 \, \epsilon(u) \right) = 
- 2 \, \nabla \left( 
\frac{\partial u}{\partial t} \, \nabla u \right)
\end{equation*} 
and, finally, the conservation law

\begin{equation} \label{energyconservation}
\frac{\partial \epsilon(u)}{\partial t} = 
\nabla \left( 
\frac{\partial u}{\partial t} \, \nabla u \right) \, \, .
\end{equation}
As is well-known,
in physical applications, $\epsilon(u)$ is called 
the energy density and 
\begin{equation*}
\frac{\partial u}{\partial t} \, \nabla u 
\end{equation*} 
is the energy flux density associated with $u$.
We note that (\ref{energyconservation}) is true also 
for $C^2$-solutions of (\ref{wavetype}) since 
for such an $u$
\begin{align*}
& \frac{\partial \epsilon(u)}{\partial t} - 
\nabla \left( 
\frac{\partial u}{\partial t} \, \nabla u \right) =
\frac{\partial}{\partial t} \, \frac{1}{2} 
\left[\left(\frac{\partial u}{\partial t}\right)^2 + 
|\nabla u|^2 + V u^2 \right] - 
\nabla \left( 
\frac{\partial u}{\partial t} \, \nabla u \right) \\
& = \frac{\partial u}{\partial t} \,
\frac{\partial^2 u}{\partial t^2} + 
(\nabla u) \cdot  \nabla \frac{\partial}{\partial t} \, u 
+ V u \, \frac{\partial u}{\partial t} -  
\left( \nabla \frac{\partial}{\partial t} \, u \right)
\cdot (\nabla u) - \frac{\partial u}{\partial t}  \, 
\Delta u \\
& = \frac{\partial u}{\partial t} \left( 
\frac{\partial^2 u}{\partial t^2} - \Delta u  + 
V u
\right) = 0 \, \, .
\end{align*} 

\section{The commutator of the wave operator and $S$}

In the following we give without proof the commutator of 
the wave operator and the new symmetry operator $S$:
 
\begin{equation} \label{commutator}
[\Box \, ,S]u =  \left( 
C_1 + C_2 \partial_r + C_3 \partial_\theta \right) \Box \, u   
\end{equation}
where 
\begin{align*}
& C_1 := \frac{2 M a^2}{\Sigma (\Delta \bar{\Sigma})^3} \left\{ 
\sin^2\!\theta \left( T_1 T_2 - \Delta \bar{\Sigma} \Delta T_3 \right)
\right.  \\
& \left. \qquad \, \, \, 
+ 2r (r^2 + a^2)^3 \left[ \Delta \bar{\Sigma} \, (2\cos^2\!\theta - \sin^2\!\theta)
+ 4 a^2 \Delta \sin^2\!\theta \cos^2\!\theta  \right] \right\}  \, \, , \\
& T_1 := (r^4 - a^4) \, \Sigma + 2 a^2  r^2 \Delta \sin^2\!\theta    \, \, , \\
& T_2 := 2\left[ 
(r - M) \Delta \bar{\Sigma}+ 2r (r^2 + a^2) \Delta - 2M(r^4 - a^4)\right] 
 \, \, , \\
& T_3 := 2 \left[ r(r^4 - a^4) + 2r^3 \Sigma + 
2 a^2 r \Delta \sin^2\!\theta + 2 a^2 r^2 (r - M) \sin^2\!\theta 
 \right]  \, \, , \\
& C_2 := - \frac{4M  a^2 \Delta \sin^2\!\theta \, [(r^4 - a^4)\Sigma + 
2 a^2 r^2 \Delta \sin^2\!\theta 
]}{\Sigma(\Delta \bar{\Sigma})^2}  \, \, , \\
& C_3 := \frac{8M a^2  r (r^2 + a^2)^3 \sin \theta \cos \theta}{\Sigma
(\Delta \bar{\Sigma})^2} 
\end{align*}
and $u \in C^4(\Omega_{r},{\mathbb{C}})$. Note that the form of the 
commutator 
(\ref{commutator}) 
generalizes that in the definition of a symmetry operator given in \cite{miller} 
to symmetry operators of the second order.

\end{document}